\documentclass[preprint,superscriptaddress,prb,amsmath,amssymb,floatfix,citeautoscript]{revtex4-1}
\usepackage[utf8]{inputenc}
\usepackage{color}
\usepackage{amsmath}
\usepackage{graphicx}
\usepackage[unicode=true]{hyperref}
\usepackage{breakurl}

\makeatletter

\providecommand{\tabularnewline}{\\}
\newcommand{\lyxdot}{.}

\@ifundefined{textcolor}{}
{%
 \definecolor{BLACK}{gray}{0}
 \definecolor{WHITE}{gray}{1}
 \definecolor{RED}{rgb}{1,0,0}
 \definecolor{GREEN}{rgb}{0,1,0}
 \definecolor{BLUE}{rgb}{0,0,1}
 \definecolor{CYAN}{cmyk}{1,0,0,0}
 \definecolor{MAGENTA}{cmyk}{0,1,0,0}
 \definecolor{YELLOW}{cmyk}{0,0,1,0}
}

\usepackage{dcolumn}
\usepackage{bm}
\usepackage{subfigure}
\usepackage{color}

\makeatother

\begin{document}

\title{Pattern formation in binary fluid mixtures induced by short-range
competing interactions}

\author{Cecilia Bores}

\affiliation{Instituto de Química Física Rocasolano, CSIC, Serrano 119, E-28006
Madrid, Spain}

\author{Enrique Lomba}

\affiliation{Instituto de Química Física Rocasolano, CSIC, Serrano 119, E-28006
Madrid, Spain}

\email{enrique.lomba@csic.es}


\affiliation{Laboratoire de Physique Théorique de la Matiére Condensée (UMR CNRS
7600), Université Pierre et Marie Curie, 4 Place Jussieu, F75252 Paris
Cedex 05, France}

\author{Aurélien Perera}

\affiliation{Laboratoire de Physique Théorique de la Matiére Condensée (UMR CNRS
7600), Université Pierre et Marie Curie, 4 Place Jussieu, F75252 Paris
Cedex 05, France}

\author{Noé G. Almarza}

\affiliation{Instituto de Química Física Rocasolano, CSIC, Serrano 119, E-28006
Madrid, Spain}

\date{\today}

\begin{abstract}
Molecular dynamics simulations and integral equation calculations
of a simple equimolar mixture of diatomic molecules and monomers interacting
via attractive and repulsive short-range potentials show the existence
of pattern formation (microheterogeneity), mostly due to depletion
forces away from the demixing region. Effective site-site potentials
extracted from the pair correlation functions using an inverse Monte
Carlo approach and an integral equation inversion procedure exhibit
the features characteristic of a short-range attractive and long-range
repulsive potential. When charges are incorporated into the model,
this becomes a coarse grained representation of a room temperature
ionic liquid, and as expected, intermediate range order becomes more
pronounced and stable. 
\end{abstract}
\maketitle

\section{Introduction}

\label{sec:introduction}

Spontaneous pattern formation is a feature present in a diverse collection
of physical, chemical and biological systems \cite{Seul1995}. In
spite of the diverse nature of these systems, the appearance of the
emerging microphases is quite similar: in 2D systems circular droplets,
stripes or ``bubbles'' occur, and in 3D
systems one may find spherical droplets, sheets or tubes. In some
cases the patterns appear as transient states due to energy or mass
fluctuations that occur in the process of spinodal decomposition,
but sometimes these states can be stabilized due to the presence of
competitive interactions, in which one of the interactions is responsible
for inhibiting the phase separation\cite{Godfrin2014,Almarza2014}.

The understanding of this self-organizing capability of soft and fluid
matter is critical for a wide panoply of applications of great relevance
nowadays. These self-assembly mechanisms play a crucial role in processes
involving protein solutions in food products \cite{Mezzenga2005,Alexander2002},
therapeutic monoclonal antibodies \cite{Yadav2012,Yearley2013,Johnston2012},
nanolithography \cite{Choi1999} or gelation processes\cite{Charbonneau2007}.

In the realm of colloidal science, systems with extremely short ranged
repulsive interactions are often used as an experimental realization
of the hard sphere fluid\cite{Royall2013}, a system notorious for
its theoretical interest. On the other hand, the addition of non-adsorbing
polymers to the colloidal solution typically activates an attractive
inter-particle interaction, due to the depletion mechanism. Moreover,
changing the concentration and molecular weight of the polymer, the
attraction range and strength of the colloid-colloid interaction can
be tuned. Clustering is to be expected due to the presence of the
attractive forces \cite{Segre2001,Poon1997}, but in principle it
would correspond to meta-stable states and/or irreversible processes
of kinetic nature. Nevertheless, microphases formed by clusters and
percolating structures can be stabilized in protein solutions and
colloid-polymer mixtures both in experiment\cite{Stradner2004} and
in theoretical descriptions\cite{Groenewold2001} due to the presence
of additional repulsive interactions stemming from electrostatic forces.
An extreme example of the stabilizing role of charges is the nanostructural
organization that appears in room temperature ionic liquids (RTIL)\cite{Canongia2006}.
In fact, it has been shown, that long range repulsive interactions
alone can give rise to nanostructural order\cite{Camp2003}, the driving
force of attractive interactions to induce spontaneous aggregation
being replaced by external forces (e.g. pressure).

In the case of colloidal systems, in which charged colloidal particles
are screened by ions in the solvent, the colloid-colloid interaction
has been shown on theoretical grounds to be adequately represented
by a Yukawa potential\cite{Derjaguin1941,dlvo_1948} according to
the Derjaguin-Landau-Verwey-Oberbeek (DLVO) theory. Following this,
numerous works have resorted to potentials with a combination of short
range attraction and a long range repulsion (SALR) in the form a double
Yukawa \cite{Bomont2012,Godfrin2013}, or a Lennard-Jones (LJ) plus
a Yukawa interaction\cite{Sciortino2004,Godfrin2014,Mani2014} in
order to model the spontaneous emergence of microstructured patterns
in fluids. On the other hand, back in 1999, Sear {\em et al.}, \cite{Sear1999}
made use of an empirical two exponential form with SALR characteristics
in order to explain the experimental appearance of stable microphases
of nanoparticles at the air-water interface. This potential has been
studied in depth in model systems, both in bulk and in confinement\cite{Imperio2004,Imperio2006,Schwanzer2010,Archer2008,Lomba2014a},
and as a rough approximation to account for vegetation patterns in
ecosystems with limited resources\cite{Meyra2014}.

In this work we will explore the possibility of pattern formation
in a system in which only short ranged forces are present. Our model
system, composed of heteronuclear dimers and monomers combines attractive
and repulsive potentials, so as to mimic the interactions present in
RTILs, but without electrostatic forces. To that aim we have performed
extensive molecular dynamics simulations in the canonical and in the
isothermal-isobaric ensembles. We will address the emergence of intermediate
range order (IRO) analyzing the behavior of the partial, and concentration-concentration
structure factors and performing a cluster analysis for various degrees
of asymmetry in the sites of the diatomic particles. Reference Interaction
Site Model (RISM) integral equation calculations have also been carried
out, and are shown to agree remarkably well with the simulations results.
By means of an Inverse Monte Carlo approach\cite{PRE200368011202},
we have extracted effective interactions from the pair correlation
functions of the simulated mixtures. For comparison, another set of
effective potentials has been obtained from the RISM results using
an integral equation inversion procedure. We will see, that despite
the fact that all interactions at play are short ranged, their net
effect leading to the pattern formation (microheterogeneity, or microstructure
segregation at the nanoscale) translates into the appearance of effective
interactions that agree with the characteristic trends of a short
range attraction and a long range repulsion, i.e. a SALR potential.
We have found that the effective potentials extracted from the simulation
and those derived by the theoretical approach agree remarkably well.
Finally, we have analyzed the role of charges on our model, which
in fact by the addition of electrostatic site-site interactions becomes
a rough representation of a RTIL. As expected, charges will be shown
to enhance the pattern formation and the stability of the nanostructured
phases.

The rest of the paper can be sketched as follows. In the next section
we introduce the model in full detail and briefly summarize the methodology.
In Section III we introduce our most significant results. Conclusions
and future prospects are to be found in Section IV.

\section{Model and methods}

\label{sec:model_methods}

Our model consists in an equimolar fluid mixture of two different species,
a two-site dimer AB, and a monomer C. The dimers are represented by
a two center Lennard-Jones (LJ) site-site potential, in which the
sites are separated by a distance $l$. Our monomers also interact
via LJ potentials. In all cases, the interactions are cut and shifted
at a distance $r_{c}$, by which the explicit form of the site-site
potentials is 
\begin{equation}
\begin{aligned}u_{\ij}(r)=4\epsilon\left[\left(\frac{\sigma_{ij}}{r}\right)^{12}-\left(\frac{\sigma_{ij}}{r}\right)^{6}-\left(\frac{\sigma_{ij}}{r_{c}}\right)^{12}+\left(\frac{\sigma_{ij}}{r_{c}}\right)^{6}\right]\;\;\mbox{if}\; r<r_{c},\end{aligned}
\end{equation}
and $u_{\ij}(r)=0$ otherwise. Our model is to a certain degree inspired
by the simple coarse-grained model for imidazolium based RTIL of Merlet
et al.\cite{Merlet2012}. We will see to what extent a simple model,
with just two sites and purely short ranged interactions can reproduce
the presence of nano-structural order as found in RTILs. To that aim
we will however preserve the attractive/repulsive character of the
interactions in the RTIL. In our model then, C monomers would correspond
to anions, AB dimers to the molecular cations, with the imidazolium
ring that contains the positive charge, being represented by site
A, and the non-polar tail, by the larger site B. This implies that
AA and CC interactions will be repulsive, BB and AC are attractive,
finally BC and AB interactions are also repulsive. For the sizes of
A and C particles we have chosen $\sigma_{AA}=\sigma_{CC}=4$ \AA \,,
the elongation of the dimer $l=8$\AA. 
The AB distances of the dimers are  fixed as  constraints of the equations
of motion.
 The LJ well is set to
$\epsilon=2.092$ kJ/mol, identical for all interactions. Since the
size of the non-polar tail is essential to determine the nanostructural
ordering\cite{Canongia2006}, we have considered various sizes for
$\sigma_{BB}$ (with $\sigma_{BB}>\sigma_{AA}$ always). For the attractive
interactions we have truncated and shifted the LJ potential at $r_{c}=3\sigma_{BB}$.
For the repulsive interactions, we have simply used $r_{c}=2^{1/6}\sigma_{ij}$,
thus defining purely repulsive soft spheres following the prescription
of Weeks, Chandler and Andersen (WCA)\cite{Weeks1971}. The complete
set of parameters for all interactions is summarized in Table \ref{table:potential}.
Finally, in order to analyze the effect of charges on the intermediate
range order, we have considered explicitly the same model with a positive
charge $+q$ on the A sites and a corresponding negative charge $-q$
on the monomers. The value of $q$ is varied between 0 and $0.25e$,
where $e$ is the elementary electron charge. Again these values are
of the same order as those considered in the model of Ref.~\onlinecite{Merlet2012}.

\subsection{Simulations and analysis}

We have carried out extensive molecular dynamics simulations of the
system previously described using the LAMMPS package\cite{LAMMPS2012,LAMMPS1995,LAMMPS2011},
in the canonical and isothermal-isobaric ensembles using a Nose-Hoover
thermostat and barostat\cite{FrenkelSmitbook}. Our samples contained
16384 particles (samples of up to 65536 particles were investigated
and no significant size dependence was found). For simplicity 
we considered equal masses for the three interaction centers: $m_A= m_B= m_c = 16$ g mol$^{-1}$. Initial thermalization
runs at a temperature of 226 K were $2\times10^{6}$ steps long, with
a time step of 1 fs. Production runs were $5\times10^{6}$ steps long,
and averages were carried out every 5000 steps.

One of the problems one can encounter when performing canonical simulations
in this type of system is the occurrence of phase transitions, either
vapor-liquid equilibria or demixing. In order to guarantee that the
states under consideration correspond to thermodynamic equilibrium
conditions, and consequently any potential intermediate range order
is not the result of a spinodal decomposition, we have run additional
isothermal-isobaric simulations and analyzed the volume fluctuation
of the samples. In this way one can avoid those states that lie inside
the liquid-vapor spinodal. Moreover, one can compute the partial
structure factors, defined as 
\begin{equation}
S_{ij}({\bf k})=x_{i}\delta_{ij}+x_{i}x_{j}\rho\int\left(g_{ij}(r)-1\right)e^{-{\bf kr}}d{\bf r}\label{skab},
\end{equation}
where $\rho$ is the total number density, $\delta_{ij}$ is a Kronecker
$\delta$, and $x_i$ is the molar fraction of component $i$. Here sites A and B are
considered as different particles and $g_{ij}$ is the atom-atom pair
distribution function. Our samples are large enough to allow for an
accurate integration of the pair distribution functions, and the results
are consistent with direct $k$-sampling. Notice that as far as Eq.(\ref{skab})
is concerned, $x_{A}=x_{B}=x_{C}=1/3$, hence in the large $k$ limit
all structure factors will tend to 1/3. From the partial structure
factors it is possible to evaluate the concentration-concentration
structure factor introduced by Bathia and Thornton \cite{Bathia1970},
for which we have defined 
\begin{equation}
S_{cc}(k)=x_{AB}^{2}S_{CC}(k)+x_{C}^{2}S_{AB-AB}(k)-2x_{AB}x_{C}S_{C-AB}(k),\label{scc}
\end{equation}
where now one has to consider explicitly the structure factors corresponding
to the molecular species AB, and as a consequence $x_{C}=x_{AB}=1/2$.
We can simply approximate $g_{AB-AB}=g_{BB}$ and $g_{C-AB}=g_{CB}$,
as if the scattering length or form factor of A sites was negligible
compared to that of B sites. This is in principle not unreasonable
given the much larger size of the B sites, but in a realistic situation
one should take explicitly into account the true scattering lengths
or form factors of sites A and B. Now one has to correct for the different
values of the molar fraction when AB is considered as a single species
and Eq.~(\ref{skab}) is used in (\ref{scc}). In this way, $\lim_{k\rightarrow\infty}S_{cc}(k)=x_{c}x_{AB}=1/4$.
With all this in mind, the presence of a divergence when $k\rightarrow0$
in $S_{cc}(k)$ is a signal of a demixing transition, so this quantity
will be essential to assess the stability of the thermodynamic states
chosen for our simulations.

Finally, back to the vapor-liquid transition, one can analyze the
corresponding $k$-dependent linear response susceptibility in density
fluctuations, namely\cite{HansenBook2nd} 
\begin{equation}
\rho k_{B}T\chi_{T}(k)=\frac{|\mathbf{S}(k)|}{\sum_{ab}(x_{a}x_{b})|\mathbf{S}(k)|_{ab}}\label{chi},
\end{equation}
whose $k=0$ limit is precisely the isothermal compressibility. In
Eq.~(\ref{chi}) $k_{B}$ is Boltzmann's constant, $T$ the absolute
temperature, and the elements of the matrix $\mathbf{S}_{ij}$ are
just the partial structure factors as defined in Eq.~(\ref{skab}).
$|\ldots|$ denotes the matrix determinant and $|\ldots|_{ab}$ the
corresponding minor of the matrix $\mathbf{S}(k)$. The presence of
a divergence --or a substantial increase in $\chi_{T}(0)$-- is a
clear indication of the vicinity of a vapor-liquid transition. A careful
monitoring of this quantity together with the use of NPT simulations
provides a reliable assessment of the stability of the state points
under consideration during the simulation runs.

All systems and conditions studied in this work are summarized in
Table \ref{table:systems}. In the case of system 8, when increasing
the charge from $0.10e$ to $0.25e$ the conditions of temperature
and density corresponding to systems 3, 6, and 7 lie in the two-phase
region. Consequently we resorted to an isothermal-isobaric simulation
at low positive pressure to achieve thermodynamic equilibrium conditions
in our system with $q=0.25e$. The final value of the total particle
density achieved in this way is indicated in Table \ref{table:systems}.

\subsection{Inverse Monte Carlo method}

With the pair correlation functions produced along the simulation
runs and the corresponding statistical uncertainties calculated using
block averages, we have used the Inverse Monte Carlo (IMC) procedure
proposed by Almarza and Lomba\cite{PRE200368011202} in order to produce
single component site-site effective potentials able to reproduce
the microscopic structure exhibited by our mixture model. The procedure
starts from a simple approximation $\beta u_{in}^{eff}(r)=-\log g(r)$
and proceeds to modify the pair potential along the simulation run
in such a way that the calculated $g^{eff}(r)$ matches the input
$g(r)$. Explicit details of the method can be found in Ref.~\onlinecite{PRE200368011202}.
In our case, we have used a total of 4000 particles.
The procedure of inversion was carried out in 20 stages. In the last stages
the effective potentials hardly  varied, and the
convergence between input and calculated
$g(r)$'s  according the prescription of Ref.~\onlinecite{PRE200368011202} was
achieved succesfully in all the cases.

In this way, one can use as input of the IMC procedure either $g_{AA}(r)$,
$g_{BB}(r)$, or $g_{CC}(r)$, and obtain a corresponding set of $u_{AA}^{eff}(r)$,
$u_{BB}^{eff}(r)$,and $u_{CC}^{eff}(r)$, which will obviously be
different, but in the case of emergence of intermediate range order
should exhibit some common features.

\subsection{RISM integral equation}

The site-site correlations are obtained by solving the usual set of
2 equations, the site-site Ornstein-Zernike equation (SSOZ) and the
closure equation, which we choose here to be the site-site hypernetted
equation (SS-HNC). The SSOZ equation for the present system is explicitly
given in the matrix form

\begin{equation}
(\mathbf{W}+\frac{\rho}{3}\mathbf{H})(\mathbf{W}^{-1}-\frac{\rho}{3}\mathbf{C})=\mathbf{I}\label{ssoz},
\end{equation}
where the $3\times3$ matrix $\mathbf{H}$ (or $\mathbf{C})$ has
for elements $H_{ij}=\tilde{h}_{ij}(k)$(or $\mathbf{C}_{ij}=\tilde{c}_{ij}(k)$),
the Fourier transform (FT) of the site-site pair correlation functions
$h_{ij}(r)=g_{ij}(r)-1$ (or the direct correlation function $c_{ij}(r)$),
where the index $i,j$ stand for one of the sites $A,B,C$. The matrix
$\mathbf{W}$ represents the intra-molecular correlations, which for
the present system gives:

\begin{equation}
\mathbf{W}=\left(\begin{array}{ccc}
\tilde{w}_{AA} & \tilde{w}_{AB} & \tilde{w}_{AC}\\
\tilde{w}_{AB} & \tilde{w}_{BB} & \tilde{w}_{BC}\\
\tilde{w}_{AC} & \tilde{w}_{BC} & \tilde{w}_{CC}
\end{array}\right)=\left(\begin{array}{ccc}
1 & j_{0}(kl) & 0\\
j_{0}(kl) & 1 & 0\\
0 & 0 & 1
\end{array}\right)\label{wkmat},
\end{equation}
where $j_{0}(x)$ is a spherical Bessel function. The matrix $\mathbf{I}$
is the identity matrix. The SS-HNC equations are written as

\begin{equation}
g_{ij}(r)=\exp\left[-\frac{u_{ij}(r)}{k_{B}T}+h_{ij}(r)-c_{ij}(r)\right]\label{sshnc},
\end{equation}
and there are 9 such independent equations to solve.

Both equations are approximate ones, and their respective inconsistencies
have been discussed many times in the past literature\cite{HansenBook2nd,Kezic2011}.
Based on empirical evidence from the literature, we expect that the
correlations obtained through these equations for the present systems,
both charged and uncharged, should be relatively good for the short
range part, but perhaps not at long range. We are particularly interested
to see if the correlations related to the appearance of the local
structures can be reproduced by this theory. The structure factor
defined in Eq.(\ref{fig:structfact}) is the appropriate function
for this purpose, as illustrated in the Results section.

The practical solution of these equations consists in discretizing
all the functions on an equidistant grid, both in $r$ and $k$ space.
We use $2048$ points with a $r$-grid of $\Delta r=0.01\sigma_{A}$,
which is enough for the present case to properly describe the asymptotic
behavior of the correlations in direct and reciprocal space. The set
of two equations are solved iteratively following techniques well
documented in the literature.

It is also possible to obtain the effective potentials which would
correspond to the equivalent one-component representation of the system.
This is achieved by imposing the  pair correlation function to be the desired
site-site correlation, namely $g(r)=g_{XX}(r)$, in the set of the
two integral equations for the 1-component system, and solve these
equations for the direct correlation function and effective pair interaction.
The direct correlation function can be obtained through the OZ equation
for 1-component system (which is  an exact relation):

\begin{equation}
(1+\rho_{S}\tilde{h}(k))(1-\rho_{S}\tilde{c}(k))=1\label{oz1comp},
\end{equation}
where $h(r)=g(r)-1=h_{XX}(r)=g_{XX}(r)-1$, and the density $\rho_{S}$
is that of the effective 1 component made solely of sites $X$.
Once $c(r)$ is obtained, one solves the HNC closure, which has the
same form as Eq.~(\ref{sshnc}), but now for the effective interaction
$u_{eff}(r)$ one gets:

\begin{equation}
u_{eff}(r)=-k_{B}T\left[\ln g_{XX}(r)+h_{XX}(r)-c(r)\right]\label{Ueff-HNC}
\end{equation}

\section{Results}

\label{sec:Results}

\subsection{Pair structure.}

\begin{figure}[h]
\subfigure[$\sigma_{BB}=8$\AA]{\includegraphics[width=0.49\textwidth]{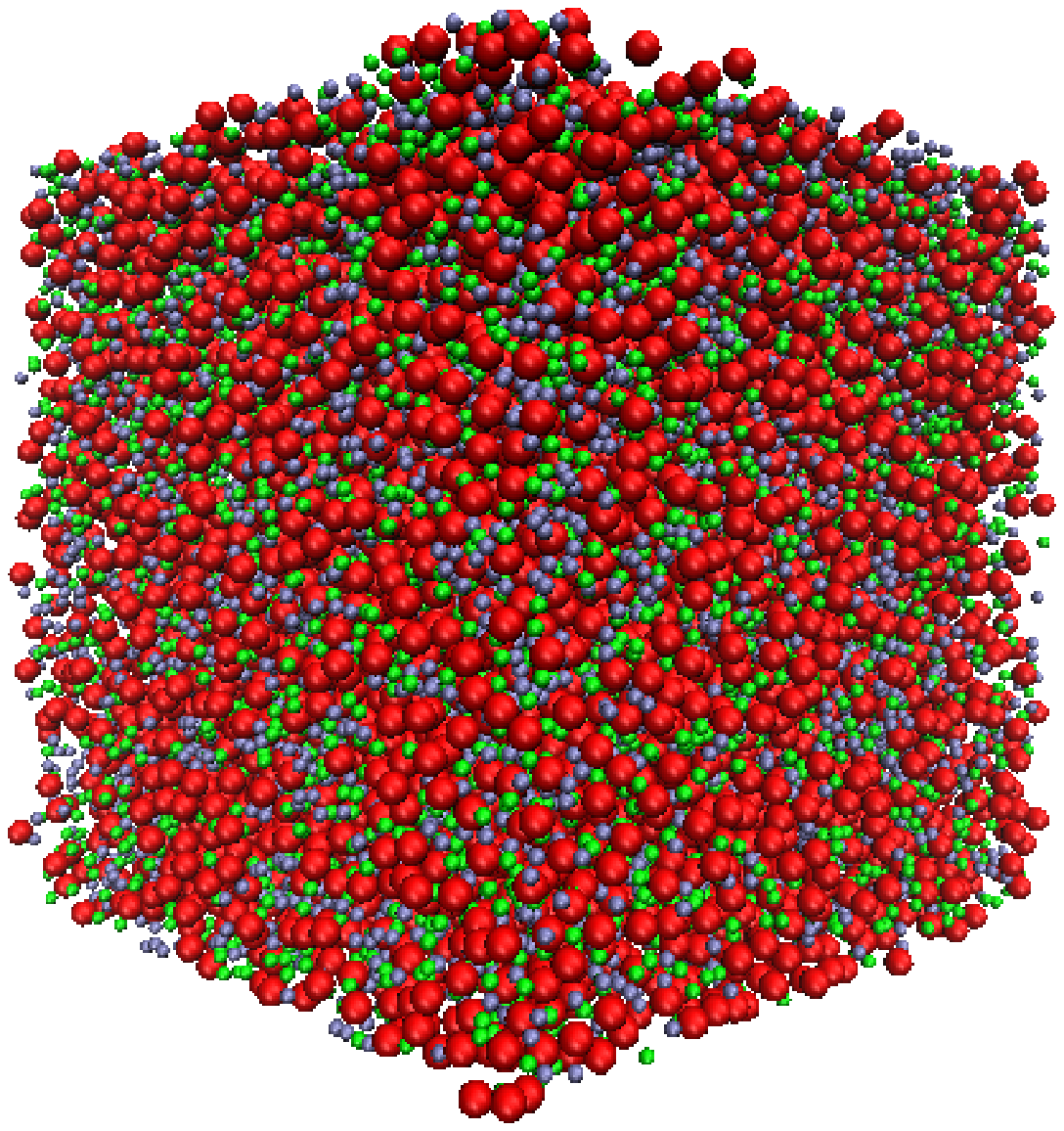}}
\subfigure[$\sigma_{BB}=12$\AA]{\includegraphics[width=0.49\textwidth]{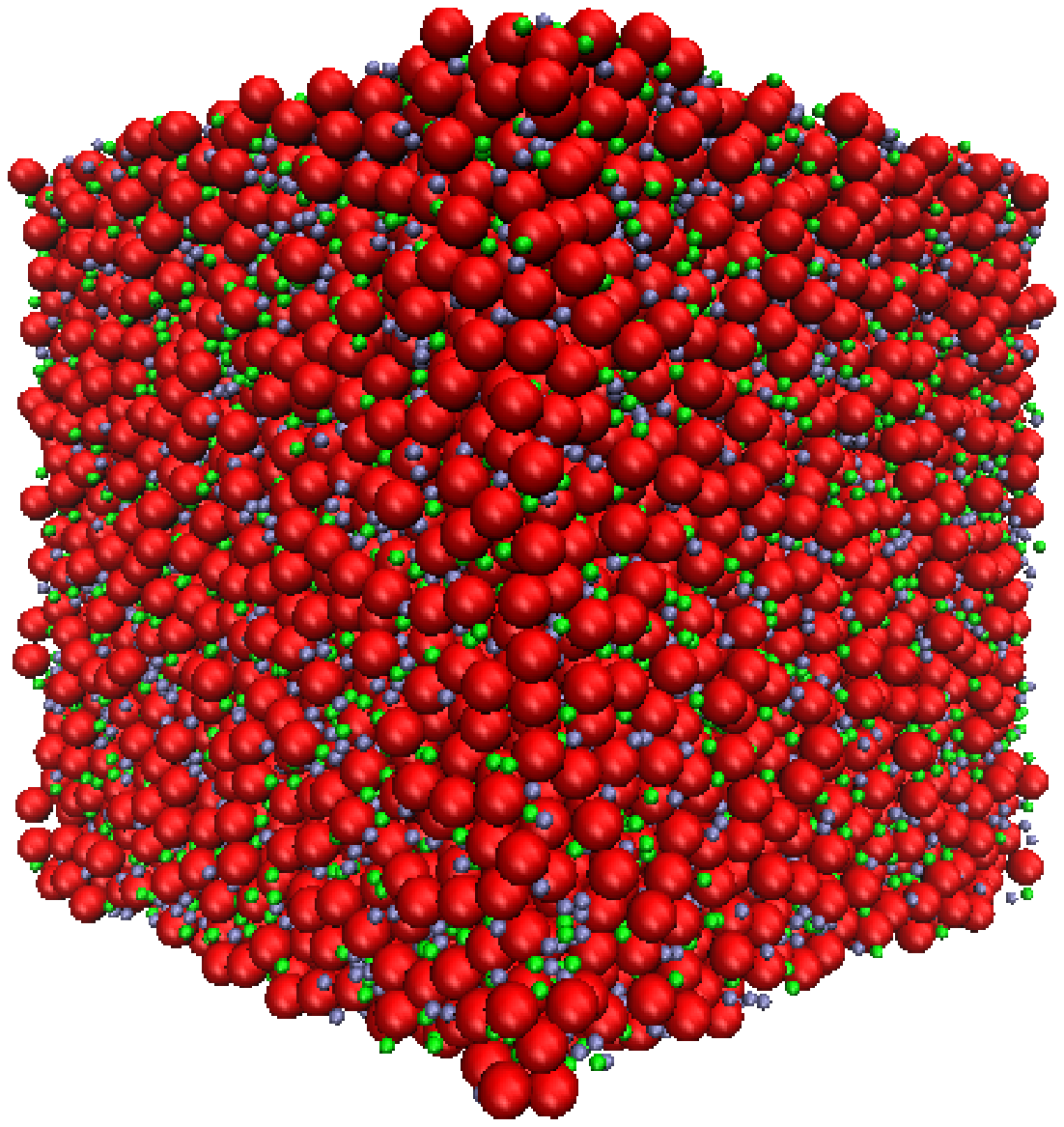}}
\protect\protect\caption{Snapshots of configurations for total particle density $\rho=0.00125$\AA $^{-3}$
and temperature $T=226.45K$ for two B-site diameters. As the size
of B-sites grows C monomers cluster in the cavities formed by the
B-sites due to excluded volume effects. All other diameters and total
density are kept fixed, $\sigma_{AA}=\sigma_{CC}=4.0$ \AA \ .}

\label{fig:bigB} 
\end{figure}

Here we have analyzed the effect of the molecular geometry on the
nanostructure formation changing the diameter of $\sigma_{BB}$. We
have first considered, $\sigma_{BB}=8$ \AA \,, 9 \AA \,, 10 \AA ,
and 12 \AA \,. Some snapshots of configurations for varying $\sigma_{BB}$
are depicted in Figure \ref{fig:bigB}. We have found that for $\sigma_{BB}>9$\AA \,
clustering or microheterogeneity of C particles can only be appreciated
when the packing of the B sites is so high that it resembles that
of a solid. In fact in this case, the height of the first peak of
$S_{BB}(k)$ exceeds 2.7, which according to the Hansen-Verlet rule\cite{PR_1969_184_151}
indicates that freezing conditions have been reached. Moreover, the
prepeak in the structure factor characteristic of the presence of
IRO is absent from $S_{BB}(k)$. The clustering of C particles results
from a merely steric effect, since these are restricted to occupy
the holes between the large B particles. These effects can be appreciated
in the snapshots of Figure \ref{fig:bigB}, where the dense packing
of B sites (red spheres) is readily apparent.

\begin{figure}[h]
\subfigure[$\sigma_{BB}=8$\AA]{\includegraphics[clip,width=0.48\textwidth]{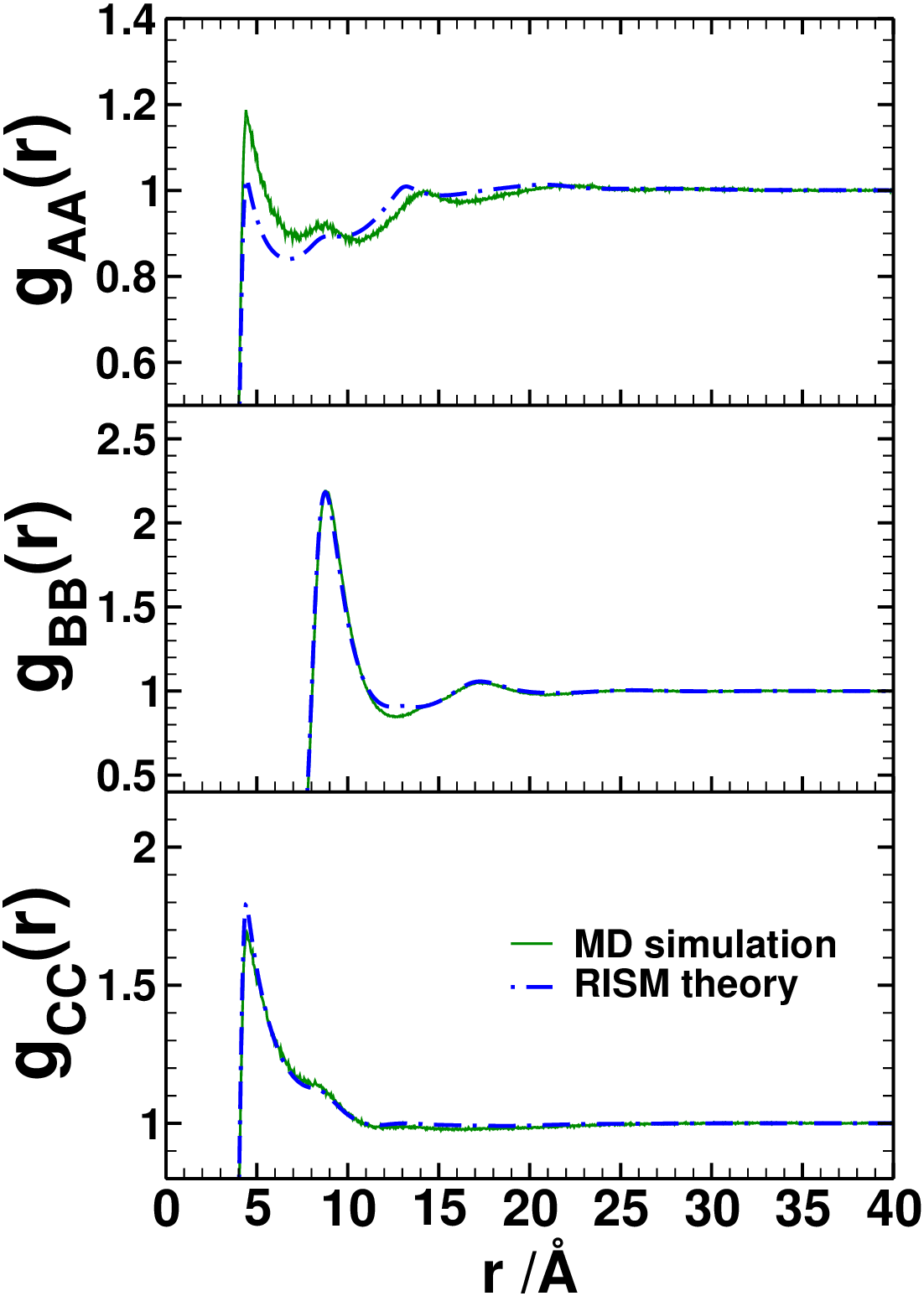}}
\subfigure[$\sigma_{BB}=9$\AA]{\includegraphics[clip,width=0.48\textwidth]{gr_rhoS_B4\lyxdot 5_T09.eps}}
\protect\protect\caption{The figures show the radial distribution functions for A, B and C
particles respectively. Column (a) corresponds to
$\sigma_{BB}=8$\AA \ for system 3 (theory vs. simulation) and column
(b) presents the simulations results for systems 4 to 6 for $\sigma_{BB}=9$\AA \,.
Total density is indicated in the legend. Simulation results are represented
by solid lines and dash-dotted curves correspond to integral equation
calculations.
}

\label{fig:pairdist} 
\end{figure}

For the reason mentioned above, we will concentrate on the results
for $\sigma_{BB}=8$ \AA \,, and $9$ \AA \,. Already in the
corresponding snapshot of Figure \ref{fig:bigB} one can appreciate
the formation of a bicontinuous network of percolating clusters, connecting
both AB dimers and C monomers. By bicontinuous network, we mean that
the clusters formed by B-sites and C particles will be seen to both
span practically the whole 
sample, forming two continuous interpenetrated percolating
microphases. This can be analyzed from a more quantitative 
perspective by first taking a look at the corresponding pair
distribution functions 
and partial structure factors, which are depicted in Figures \ref{fig:pairdist}
and \ref{fig:structfact} respectively for Systems 1 to 6. Focusing
first on the $g_{AA}$ pair distribution function, one first appreciates
the large exclusion hole after the first layer, which is a simple
consequence of the large size of B-sites. Obviously the exclusion
hole grows with the size of the B-sites, as can be seen when comparing
Figures on the left and right columns. Correlations between A-sites
extend up to five $\sigma_{AA}$, and the width of the $g_{CC}$ correlation
is $\approx2\sigma_{CC}$. These features hint at the presence of
some degree of IRO. B-B correlations (graphs in the middle row) behave
like those of a dense fluid, and no apparent sign of clustering or
IRO is evident. In contrast, the wide first peak of $g_{CC}$ is characteristic
of clusters of particles confined in cavities, in this case formed
by $B$-sites. This effect, as mentioned before is maximized for the
largest $\sigma_{BB}$. We will see later, that these clusters of
partly occluded C-particles are connected, forming a three dimensional
percolating structure.

If we take now a look at the partial structure factors, we immediately
appreciate a feature characteristic of the emergence of IRO, namely
the presence of a prepeak at 0.25\AA $^{-1}$. This corresponds to
correlations in the range of 25\AA \,, the distance at which any sign
of structure of the pair distribution function dies out. Interestingly,
the prepeak is almost absent in $S_{AA}$, except for a small maximum
visible for the $\sigma_{BB}=9$ \AA \, and the highest density.
This quantity shows otherwise very little structure for $k>0.5$ \AA $^{-1}$.
As seen in the $g_{AA}$'s, the most relevant feature in the AA correlations
is the exclusion hole due to the presence of the $B$-sites. In contrast,
$S_{BB}$ does exhibit a prepeak, even when no evidence of IRO was
visible in $g_{BB}$. This prepeak is more apparent in the monomer
structure factor $S_{CC}$. When the density is lowered the prepeak
in the B-site structure factor shifts to lower $k$-values, and vanishes
at $\rho=0.001$\AA $^{-3}$. In the case of $S_{CC}$, the position
of the prepeak is preserved, but its magnitude decreases. In Figure
\ref{fig:conconS} the corresponding concentration-concentration structure
factor is displayed. The prepeak at $k_{0}\approx0.25$\AA $^{-1}$
is preserved, although its magnitude decreases when the total density
is lowered. In contrast no increase when $k\rightarrow0$ is visible.
This implies that we are encountering concentration fluctuations inducing
spatial inhomogeneities, but no demixing transition. In Figure \ref{fig:isocomp}
we have plotted the k-dependent susceptibility corresponding to density
fluctuations. The prepeak is visible except for the lowest density,
which implies that density inhomogeneities with a spatial patterns
are also correlated with the corresponding concentration inhomogeneities.
But now, the $k\rightarrow0$ behavior is different. As density is
decreased the susceptibility (i.e. the isothermal compressibility)
grows, an indication of the vicinity of a vapor-liquid transition.
This means, that lowering the density from the value of $\rho=0.001$\AA $^{-1}$
at the same temperature could move the system across the spinodal
curve into the two-phase region. Our analysis indicates that the thermodynamic
conditions we have simulated can be considered equilibrium states.
Moreover, we have confirmed that the results do not have a significant
sample size dependence, by which metastability can also be ruled out.

The site-site correlation functions and structure factors obtained
from the RISM theory are represented in dashed lines in Figs.2-3.
It is seen that the agreement is excellent in most cases, particularly
in what concerns the BB and CC correlations. The AA correlations are
systematically underestimated near contact and overestimated at larger
distances. The most significant differences are seen for the structure
factors in Fig.3. Integral equations tend to exaggerate concentration
fluctuations, and often tend to interpret small aggregate formations
as such\cite{Perera2013,Perera2015} . We observe here a similar
trend for the low density case $\rho=0.001$\AA $^{-3}$, for which
fluctuations compete the most with aggregate formation. The prediction
of aggregation, through the pre-peak is in very good agreement with
simulations for the highest density $\rho=0.0015$\AA $^{-3}$, precisely
when the denser packing tends to favor aggregation. This is also in
line with previous observations of similar type of behavior for model
ionic liquids. These features are a direct consequence of the fact
that the HNC closure approximation misses high order correlations,
hence high order cluster contributions, which are represented in the
bridge term $b_{ij}(r)$ that is neglected in the exponential of Eq.~(\ref{sshnc}).
We observe that in all cases, the k=0 behavior of the RISM structure
factor always overestimates the concentration fluctuations.

\subsection{Effective pair potentials}

In Figure \ref{fig:effpot} we present the effective potentials obtained
from the site-site pair distribution functions. By construction, using
these effective potentials in a simulation for a single component
system will lead to a pair distribution function coincident with the
original site-site correlation of the mixture. This is one of the
possible alternatives to reduce the behavior of a complex system to
a simpler one component system. Other alternatives, such as the force-matching
approach\cite{JCP_2004_120_10896} will lead to quantitatively different
results, but certainly retaining the essential features of the effective
potentials found here. Among these features, we see that in all cases
the effective potential has a short range (extremely short in the
case of AA potentials) attractive well and this is followed by a long
range repulsive region, which extends to 20-30 \AA \,. The repulsive
region of $U_{CC}^{eff}$ is much less visible and is illustrated
in the inset. The repulsive range is more influenced by the change
in the total density. The attractive part of AA and CC effective interactions
is due to depletion forces (in this case the plain site-site interactions
are repulsive). In the case of AA interactions, most of the attractive
well is masked by the excluded volume effect of the B sites in the
AB molecules (the large repulsive potential between 5-15 \AA \ corresponds
to the exclusion hole in $g_{AA}$). Note that even if in $g_{BB}$
long range correlations due to nanostructure organization are clearly
not visible, there are long range repulsions in the BB effective potential,
which are reflected in the prepeak in $S_{BB}$ as an indication of
IRO. The long range repulsion vanishes for $\rho=0.001$\AA $^{-3}$,
which we have seen is a state approaching the gas-liquid transition.

Fig.~\ref{fig:effpot} shows the effective pair potentials as obtained
by the integral equation approach outlined in Section C. the comparison
with the simulations is overall quite good in all cases. However,
it is seen that the repulsive shoulder -which is the signature of
the clustering ability- is always systematically underestimated by
the theory. This is a direct consequence of the weaker tendency of
the IET to predict clustering.

Taken into account that B-sites are much larger that A-sites, we can
think of our model as a system of B particles in a ``sea'' of C
monomers, just like colloids in solution. Following Mani et al.\cite{Mani2014}
we can use a functional form of the type 
\begin{equation}
\begin{aligned}U(r)/k_BT=4a_{0}\left[\left(\frac{\sigma_{BB}}{r}\right)^{12}-\left(\frac{\sigma_{BB}}{r}\right)^{a_{1}}\right]+\frac{a_{2}a_{3}}{r}e^{{-\frac{r}{a_{3}}}}\label{fit-eq}\end{aligned}
\end{equation}
to represent the BB effective interactions. 
Note that given the large
size of the B-sites, we have retained the repulsive part of the bare
LJ interaction in order to account for the repulsive component of
the effective potential. One can see that the fits of the effective
interactions $U_{BB}^{eff}/(k_{B}T)$ to Eq.~(\ref{fit-eq}) represented
in Figure \ref{fig:fiteffpot} are fairly accurate except for the
minor inflection of the curve around 13 \AA \,. The parameters of
the fit are collected in Table \ref{table:effpotpar}. Notice that
the exponent of the attractive LJ component, $a_{1}$ deviates substantially
from the standard value of 6, being its range shorter as density increases.
The range parameter $a_{3}$ grows considerably with the density,
reflecting the increase of intermediate range ordering as the total
density is increased. We observe that a single component representation
of our system can be well performed by a standard SALR potential in
which the long range repulsion has the form a Yukawa interaction,
even when the original bare interactions in the mixture are relatively
short ranged LJ potential.

\subsection{Cluster analysis}

In order to go beyond the mere qualitative information provided by
simulation snapshots and the two-body level information furnished
by pair distribution functions or site-site structure factors, we
have also performed a geometric cluster analysis on the B sites and
the C monomers, using different values for the link distance $r_{cl}$.
Essentially this distance defines two particles as linked, and in
this work it has been defined in terms of the position of the inflection
point of the corresponding effective potentials depicted in Figure
\ref{fig:effpot}. We will use various values of $r_{cl}$ in the
range 10-12 \AA \,, for B-sites and C monomers, and 6-8 \AA\, for
A-sites. The  effects of the particular choice of $r_{cl}$
on the cluster distribution will be analyzed. Specifically, we have calculated the
normalized cluster size distribution $N(s)$, as proposed by Stauffer
\cite{Stauffer1979}. This quantity is defined as the fraction of
particles contained in clusters of size $s$, i.e. $N(s)=n(s)(s/N)$, where $n(s)$ is the number of clusters of 
size $s$.  With this definition, $\sum N(s)=1$.
Of all the systems analyzed, in Figure \ref{fig:cluster} we have
chosen to plot the results of System 6, which exhibits a significant
prepeak in its partial structure factors. We observe that the normalized cluster
size distributions of both A, and, B-sites and C monomers present
the same qualitative features: first one finds a maximum for isolated
particles which decays monotonously to zero at a value of cluster
size, $s$, that strongly depends on $r_{cl}$. This is a typical
behavior of a non-associating fluid, in which instantaneous clusters
are created and destroyed as particles explore their configurational
space. If stable finite clusters were formed, the cluster size distribution
should exhibit the corresponding maxima for the preferred sizes. On
the other end of the $s$-axis, interestingly one finds large clusters
that span all the simulation cell. Here $N(s)$ shows little dependence on $r_{cl}$, particularly for the
B-sites and C monomers. Finally, the cluster size distribution
of A and B sites is qualitatively very similar, which is understandable
taking into account that both sites are linked into single molecular
units. In the next section we will see that this symmetry is broken
by the presence of charges and a new symmetry between A-sites and
C particles emerges.

Thus from our analysis, a more clear picture shows up, in which we
have a large portion of the sample linked into microsegregated clusters
forming bicontinuous structures, with a remnant of disconnected particles
that form short lived structures up to tens or hundreds of particles
depending on the choice of $r_{cl}$, as one would expect in a non-associating
fluid..

\subsection{The effect of charges}

Our previous results have shown that microheterogeneity, or stable
intermediate range order can be induced by competing short range interactions
in a simple mixture model of dimers and monomers. Our model was somehow
inspired by a coarse grained representation of ionic liquids, which
are in reality characterized by the presence of Coulombic interactions,
absent from our model. An immediate question that deserves to be answered
is then, how would the presence of charges affect the stability of
the aforementioned bicontinuous structures ? To that aim we have carried
out the corresponding analysis on systems 7 and 8, that, as mentioned,
correspond to system 3 with charges $+q$ added to sites A and $-q$
to the C monomers. For $q=0.1e$, standard canonical molecular dynamics
simulations were run. Recall that in the case of $q=0.25e$, density
had to be increased in order to move out of the vapor-liquid coexistence
region. This was simply achieved by means of an isothermal-isobaric
simulation run at the same T as the original system and a pressure
of 0.61 MPa, leading to a total $\rho=0.00195$ \AA $^{-3}$. In the
snapshots of Figure \ref{fig:charges} one can readily see that the
charges enhance the formation of microstructural order, and particularly
for the highest charge one see very well defined stripes of C particles,
stripes that now appear to be finite. A more clear picture emerges
when taking a look at the partial structure factors, presented in
left panels of Figure \ref{fig:chargesClus}. Now the prepeak is perfectly
defined even for the $S_{AA}$ structure factor for the lowest charge,
in contrast with the uncharged system $S_{AA}$. The extremely large
values of $S_{\alpha\beta}(k_{0})$ for $k_{0}\approx0.25$ \AA $^{-1}$,
resemble Bragg peaks, and indicate the presence of quasi-periodic
order in the microstructural domains. Moreover, if now one looks at
the cluster size distributions plotted on the right panels of Figure
\ref{fig:chargesClus}, together with the percolating clusters, one
finds now a maximum centered at $s\approx20$ for $q=0.25e$ for C
and A-sites, which indicates the presence of finite clusters of monomers
and A-sites. This maximum is preserved in the results obtained for
other charges up to $q=0.2e$ (not shown for the sake of brevity), to disappear
for weaker Coulombic interactions. It is obvious that the net effect
of charges on the microstructuring of our model mixture is to enhance
the formation of nanostructures, also giving rise to the formation
of finite size clusters for sufficiently high charges. In contrast,
B-sites form a percolating bicontinuous structure coexisting with
some disconnected B-sites or short lived aggregates. A-sites and C
monomer  form aggregates embedded in the percolating
network of B-sites. All this suggests that the network of B-sites
forms cavities, with the A-sites pointing inside the cavity. This
in turn is filled by C monomers. This configuration is favored both
by steric effects and by the net attraction between the positively
charge A sites and negatively charged C monomers.

On the other hand, despite the fact that A-sites form part of the
AB dimers and C monomers are independent particles, due to the symmetry
of the electrostatic interactions and the symmetry in shape and density
--$\sigma_{AA}=\sigma_{CC}$, $\rho_{A}=\rho_{C}$--, as the charges
increase, AA and CC correlations become extremely similar --compare
$S_{AA}$ and $S_{CC}$ in Figure \ref{fig:chargesClus}--, as one
would encounter in a simple fully symmetric electrolyte.

The next question is how this is all reflected on the effective potentials.
These are plotted on Figure \ref{fig:chargesEff}. In all cases one
observes the characteristic SALR structure, obviously being the CC
and AA effective interactions those that are most affected by the
introduction of charges. In spite of the fact that these two effective
interactions result from the coarse graining of many body effects,
the dominant role of electrostatic interactions already reflected
in the partial structure factors leads to surprisingly similar effective
potentials when charges are present. On the other hand the changes
in $U_{BB}^{eff}$ are just quantitative. The attractive part is hardly
influenced by the charges, since it results mostly from the depletion
interactions and the bare attractive $u_{BB}$. The long range repulsion
is enhanced, and as the charge reaches $q=0.25e$ oscillations appear.
These oscillations recall the Friedel oscillations characteristic
of effective cation-cation potentials in liquid metals\cite{Cusack1987}.
In the latter instance, the oscillations result from the quantum nature
of the electrons. Here they result from the interplay of the Coulombic
interactions and depletion forces. Thus for sufficiently large charges
the long range attractive interaction between C, and A sites propagates
through the AB bonds and induces the attraction well around 30\AA \,
as a result of a many body effect.


\section{Conclusions}

\label{sec:Conclusions}

In summary, we have shown that a simple mixture of heteronuclear AB
dimers and C monomers, with short range attractive and repulsive interactions
designed so as to mimic the interactions present in RTILs, can give
rise to the presence of nanostructural order in the form of micro-segregation
in bicontinuous structures. This in turn translates into the characteristic
presence of a prepeak in the site-site structure factors. These features
are found both in simulation and in the integral equation results.
The effective site-site potentials extracted from the pair distribution
functions by means of an IMC  and integral equation approach, display
the characteristic features of the SALR interactions, with the repulsive
long range increasing as the total density (and hence the ordering)
increases. The addition of charges to the model enhances the nanostructural
order. When charges are large enough, one finds well structured phases
in which bicontinuous structures coexist with finite size aggregates
of monomers, caged in cavities formed by a network of the large uncharged
sites, and with the cationic sites facing the inner part of the cavity.
The effect of charges on our simple and rather symmetric model induces
the symmetrization of the correlations of the anionic monomers and
the cationic sites. The microscopic structure formed by the uncharged
sites (apolar head in the RTILs) retains its bicontinuous nature and
even if it is stabilized and enhanced by the charges is still mostly
dominated by depletion effects and the bare short range attraction
of the B-sites. In this regard, it is interesting
to note that the appearance of a pre-peak in the wide angle scattering
experiments and computer simulations of RTILS have been a subject of
much investigations \cite{Annapureddy2010,Wang2007} and has been related to
the charge ordering and the subsequent appearance of segregated charged
and uncharged molecular domains. Our work presents a unified view
of microsegregated bi-continuous domains, pre-peaks in structure factors
and SARL type interactions, which are common to many complex systems.

Obviously a much richer variety of structures would result from longer
attractive uncharged tails, beyond the single B-site model used here.
On the other hand, our simple model when reduced to two dimensions
most likely will also give rise to more complex structures, which
in three dimensions are hindered by entropic effects. This is certainly
a problem relevant to the behavior at interfaces which we intend to
address in the future.

\begin{acknowledgments}
C.B.Q., N.G.A., and E.L. acknowledge financial support from the Dirección
General de Investigación Científica y Técnica under Grant No. FIS2013-47350-C5-4-R.
E.L. gratefully acknowledges the CNRS support of his stay at the LPTMC
at the Université Pierre et Marie Curie, where most of this work was
conceived. 
\end{acknowledgments}



%

\newpage{}

\begin{table}[h]
\protect\protect\caption{Lennard-Jones potential parameters.}

\label{table:potential}

\begin{tabular}{cccccccc}
\hline 
Particle $i$  & Particle $j$  & Interaction  & $\epsilon$ (kJ/mol)  &  $\sigma_{ij}$ & &$r_{c}$  \tabularnewline
\hline 
A  & A  & repulsive  & 2.092  & 4.0 \AA  &&  $2^{1/6}\cdot\sigma_{AA}$ \tabularnewline
A  & B  & repulsive  & 2.092  &  $(\sigma_{AA}+\sigma_{BB})/2$  & & $2^{1/6}\cdot\sigma_{AB}$ \tabularnewline
A  & C  & attractive  & 2.092  & 4.0 \AA & &$3\cdot\sigma_{BB}$ \tabularnewline
B  & B  & attractive  & 2.092  & $\sigma_{BB}$  &  &$3\cdot\sigma_{BB}$ \tabularnewline
B  & C  & repulsive  & 2.092  & $(\sigma_{BB}+\sigma_{CC})/2$  & &$2^{1/6}\cdot\sigma_{BC}$ \tabularnewline
C  & C  & repulsive  & 2.092  & 4.0 \AA & &$2^{1/6}\cdot\sigma_{CC}$ \tabularnewline
\hline 
\end{tabular}
\end{table}

\begin{table}[h]
\protect\protect\caption{Potential parameters and thermodynamic state variables for the systems under study.}

\label{table:systems} %
\begin{tabular}{lccccc}
\hline 
 & \multicolumn{2}{c}{ \textbf{Potential}} & \multicolumn{2}{c}{ \textbf{Thermodynamic state }} & \tabularnewline
\hline 
 & $|q|(e)$  & $\sigma_{B}$ (\AA )  & $\rho$(\AA $^{-3}$)  & T(K)  & P(MPa) \tabularnewline
\hline 
System 1  & 0  & 8.0  & 0.001  & 226.4  & 27.05 \tabularnewline
System 2  & 0  & 8.0  & 0.00125  & 226.5  & 39.5 \tabularnewline
System 3  & 0  & 8.0  & 0.0015  & 226.5  & 59.4 \tabularnewline
\hline 
System 4  & 0  & 9.0  & 0.001  & 226.5  & 30.4 \tabularnewline
System 5  & 0  & 9.0  & 0.00125  & 226.4  & 53.2 \tabularnewline
System 6  & 0  & 9.0  & 0.0015  & 226.4  & 96.7 \tabularnewline
\hline 
System 7  & 0.1  & 8.0  & 0.0015  & 226.4  & 39.4 \tabularnewline
System 8  & 0.25  & 8.0  & 0.00195  & 226.3  & 0.61 \tabularnewline
\hline 
\end{tabular}
\end{table}

\begin{table}[h]
\protect\protect\caption{Parameters of tha SALR effective interaction (\ref{fit-eq}) between
B sites fitted to the data extracted from the IMC procedure. Note that
the potential is scaled with $k_BT$, by which $a_0$ is dimensionless.
}

\label{table:effpotpar}

\begin{tabular}{ccccc}
\hline 
 & $a_{0}$  & $a_{1}$  & $a_{2}$(\AA)  & $a_{3}$(\AA ) \tabularnewline
\hline 
System 1  & 1.788  & 8.185  & 3.749  & 4.297 \tabularnewline
System 2  & 2.066  & 8.667  & 0.843  & 7.282 \tabularnewline
System 3  & 3.578  & 9.927  & 0.231  & 14.816 \tabularnewline
\hline 
\end{tabular}
\end{table}

\begin{figure}[h]
\subfigure[$\sigma_{BB}=8$\AA]{\includegraphics[clip,width=0.48\textwidth]{Sk_rhoS_B4_T09.eps}}
\subfigure[$\sigma_{BB}=9$\AA]{\includegraphics[clip,width=0.48\textwidth]{Sk_rhoS_B4\lyxdot 5_T09.eps}}

\protect\protect\caption{The figures show the structure factors for A,
  B and C particles respectively.  Column (a) corresponds to
$\sigma_{BB}=8$\AA \ for system 3 (theory vs. simulation) and column
(b) presents the simulations results for systems 4 to 6 for $\sigma_{BB}=9$\AA \,.
Total density is indicated in the legend. Simulation results are represented
by solid lines and dash-dotted curves correspond to integral equation
calculations.
}

\label{fig:structfact} 
\end{figure}

\begin{figure}[h]
\includegraphics[width=1\textwidth]{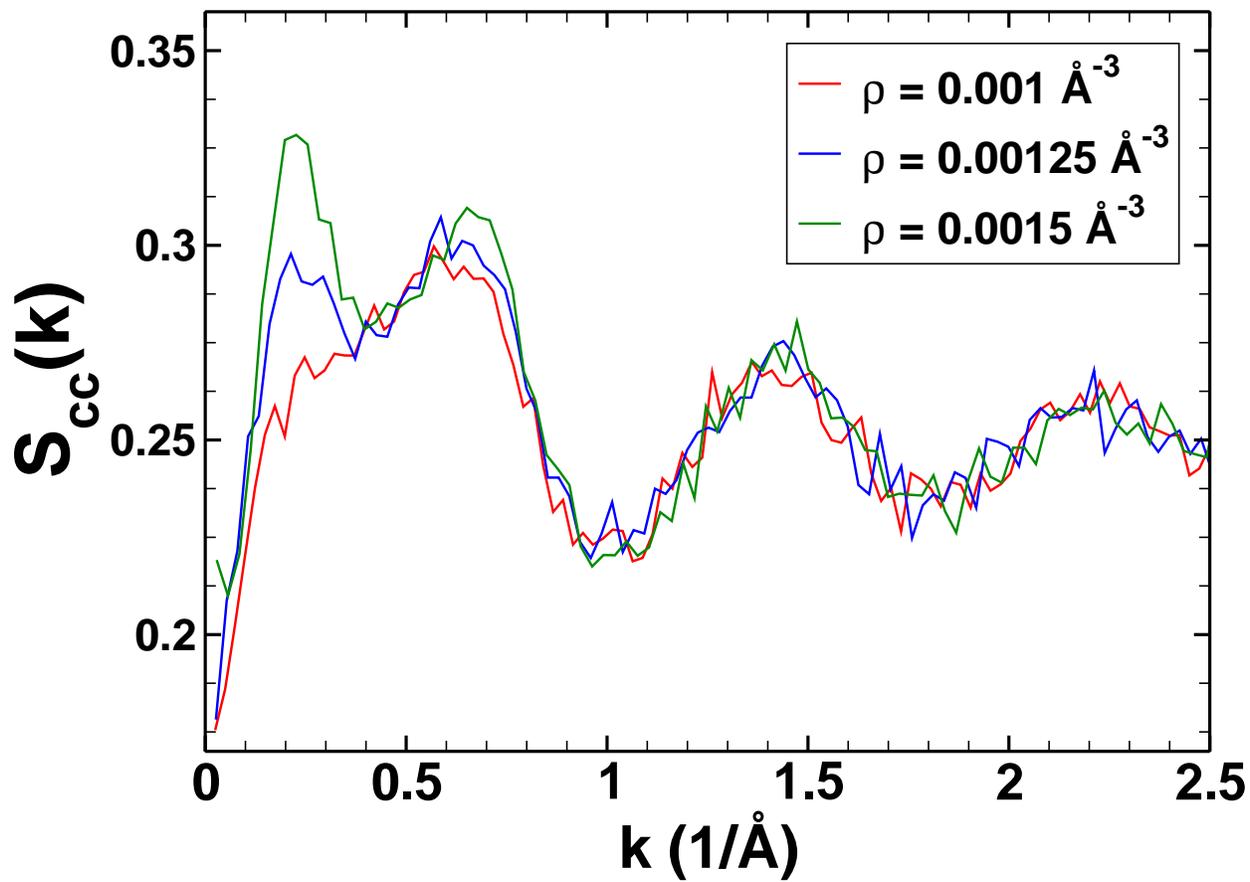}

\protect\protect\caption{Concentration-concentration structure factor for the Systems 1, 2
and 3. }

\label{fig:conconS} 
\end{figure}

\begin{figure}[h]
\includegraphics[width=0.8\textwidth]{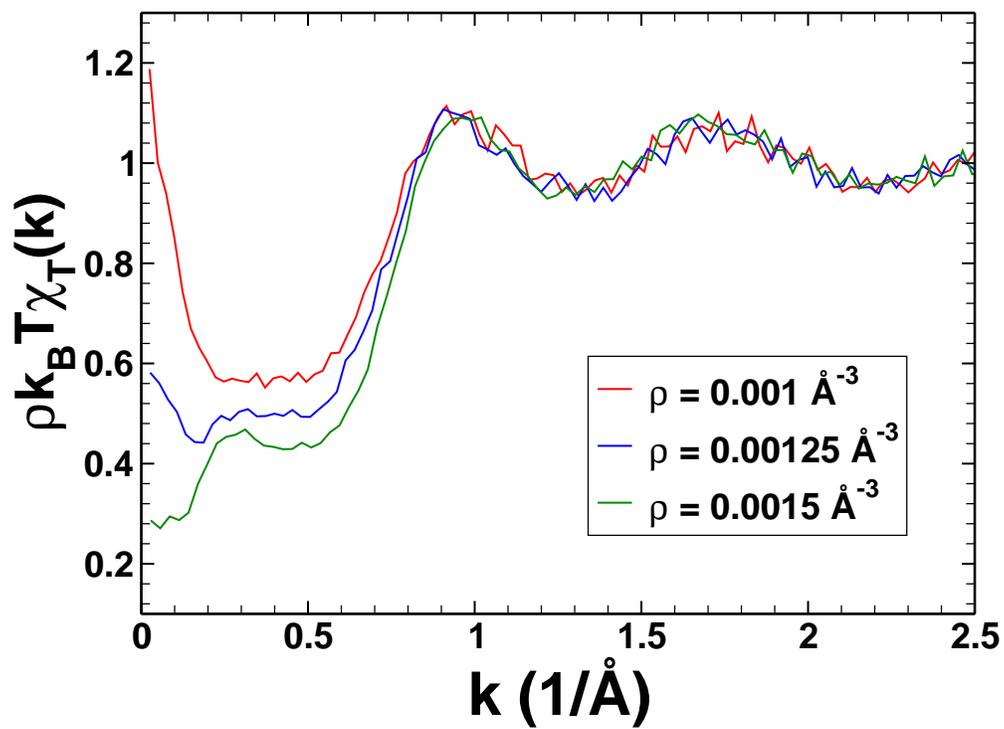}

\protect\protect\caption{Isothermal compressibility as a function of $k$ for the Systems 1,
2 and 3. }

\label{fig:isocomp} 
\end{figure}

\begin{figure}[h]
\subfigure[$\sigma_{BB}=8$\AA]{\includegraphics[clip,width=0.48\textwidth]{Effpot_rhoS_B4_T09.eps}}
\subfigure[$\sigma_{BB}=9$\AA]{\includegraphics[clip,width=0.48\textwidth]{Effpot_rhoS_B45_T09.eps}}

\protect\protect\caption{Effective potentials for A, B and C particles
  respectively.   Column (a) corresponds to
$\sigma_{BB}=8$\AA \ for system 3 (theory vs. simulation) and column
(b) presents the simulations results for systems 4 to 6 for $\sigma_{BB}=9$\AA \,.
Total density is indicated in the legend. Simulation results are represented
by solid lines and dash-dotted curves correspond to integral equation
calculations.}

\label{fig:effpot} 
\end{figure}

\begin{figure}[h]
\includegraphics[clip,width=0.65\textwidth]{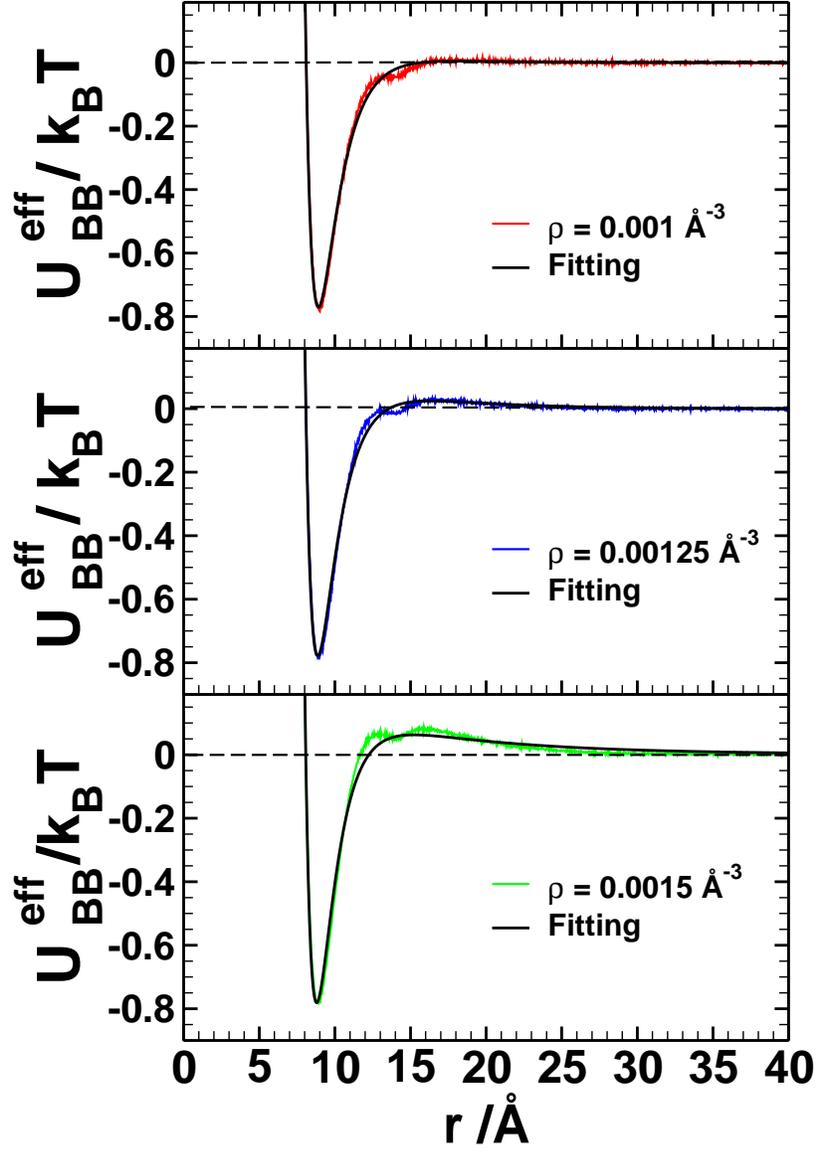}

\protect\protect\caption{B-B effective interaction for systems 1 to 3, fitted to a generalized
LJ+Yukawa interaction}

\label{fig:fiteffpot} 
\end{figure}

\begin{figure}[h]
\includegraphics[clip,width=0.8\textwidth]{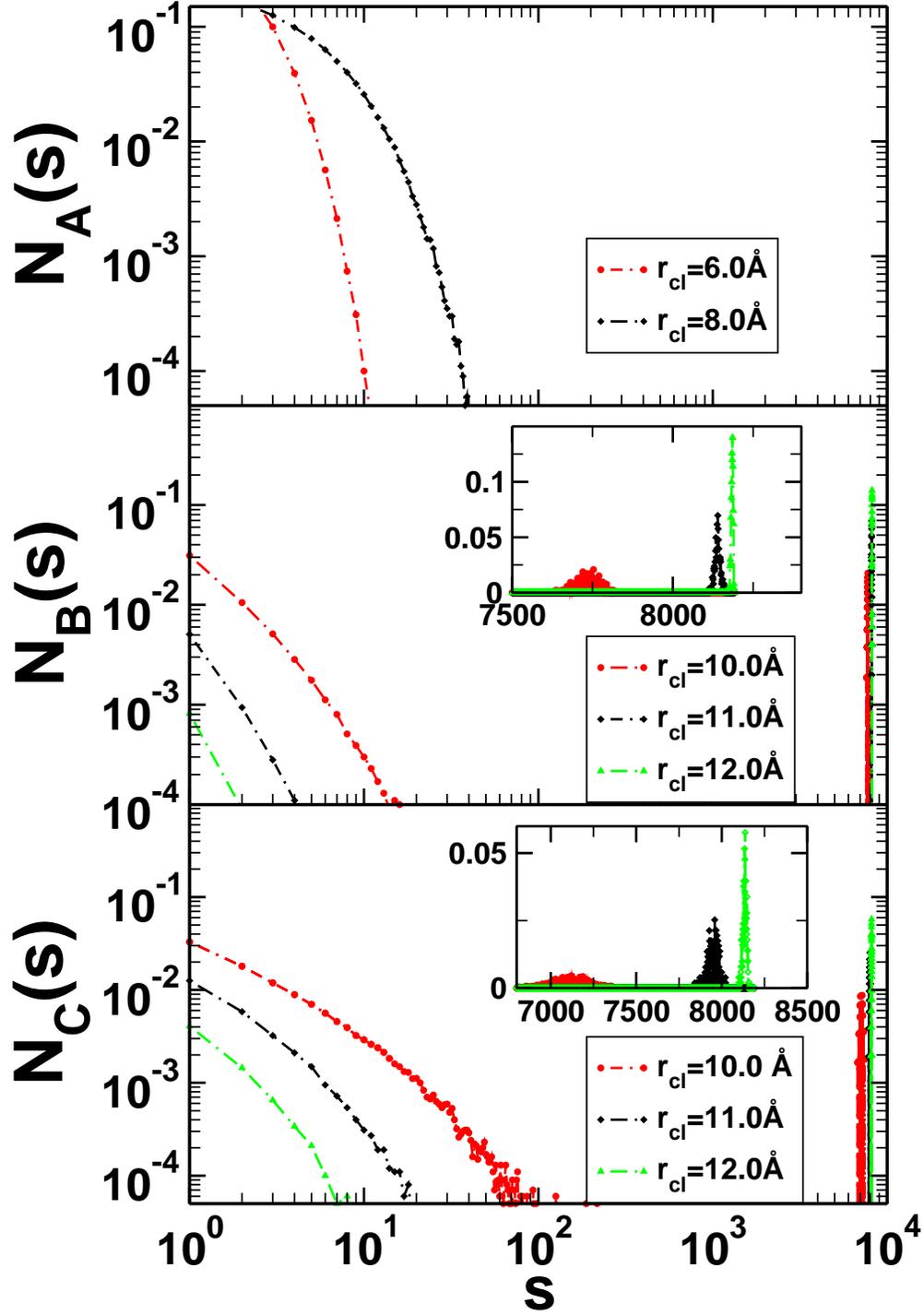}
\protect\protect\caption{Normalized cluster distribution function for A, and B sites, and C
monomers of System 3.}

\label{fig:cluster} 
\end{figure}

\begin{figure}[h]
\subfigure[$q_A=0.10e$; $q_C=-0.10e$]{\includegraphics[width=0.5\textwidth]{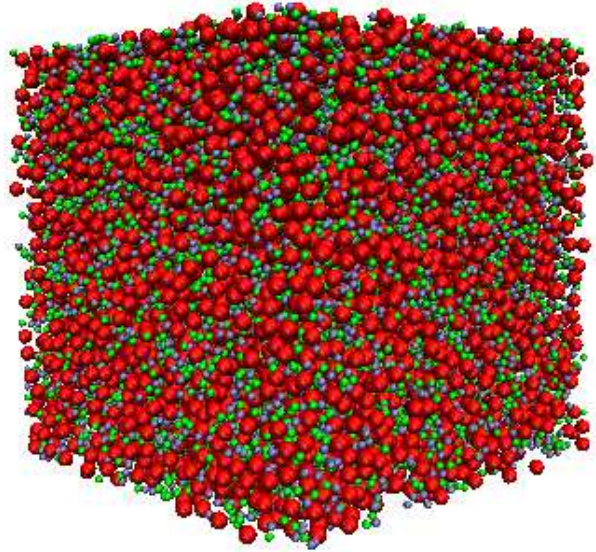}}
\subfigure[$q_A=0.25e$; $q_C=-0.25e$]{\includegraphics[width=0.5\textwidth]{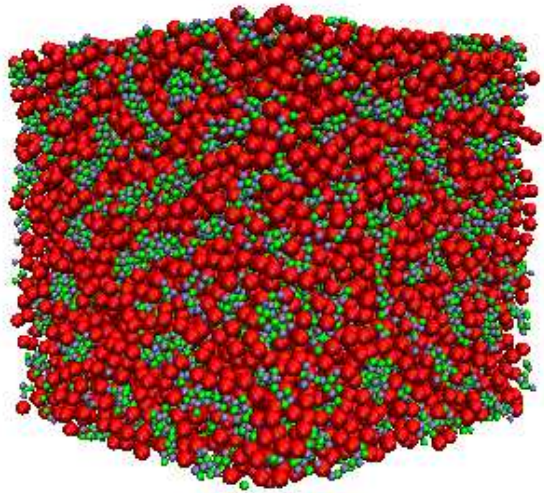}
} \protect\protect\caption{Snapshots of the equimolar mixture of AB dimers and C monomers with
embedded charges (indicated on the figures).}

\label{fig:charges} 
\end{figure}

\begin{figure}[h]
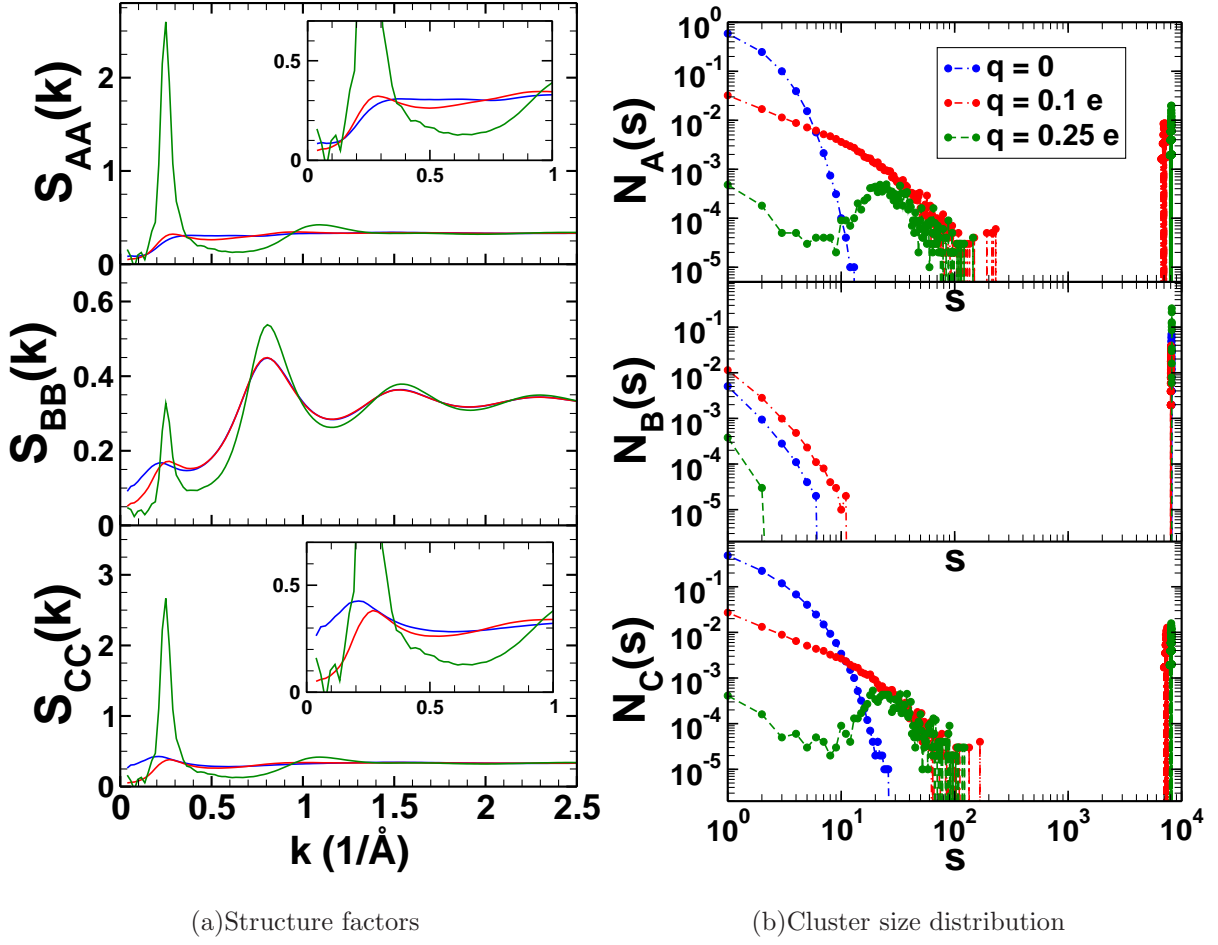

\subfigure[Structure factors]{\includegraphics[clip,width=0.48\textwidth]{Sk_rho0012_B4_T09_charges.eps}}
\subfigure[Cluster size distribution]{\includegraphics[clip,width=0.48\textwidth]{clust_rho0012_B4_charges.eps}}

\protect\protect\caption{(a) Charge dependence of the partial structure factors for A (top),
B(middle) and C (bottom) particles (b) Charge dependence of the cluster
size distribution. Charge magnitudes are specified in the legend.}

\label{fig:chargesClus} 
\end{figure}

\begin{figure}[h]
\includegraphics[clip,width=0.8\textwidth]{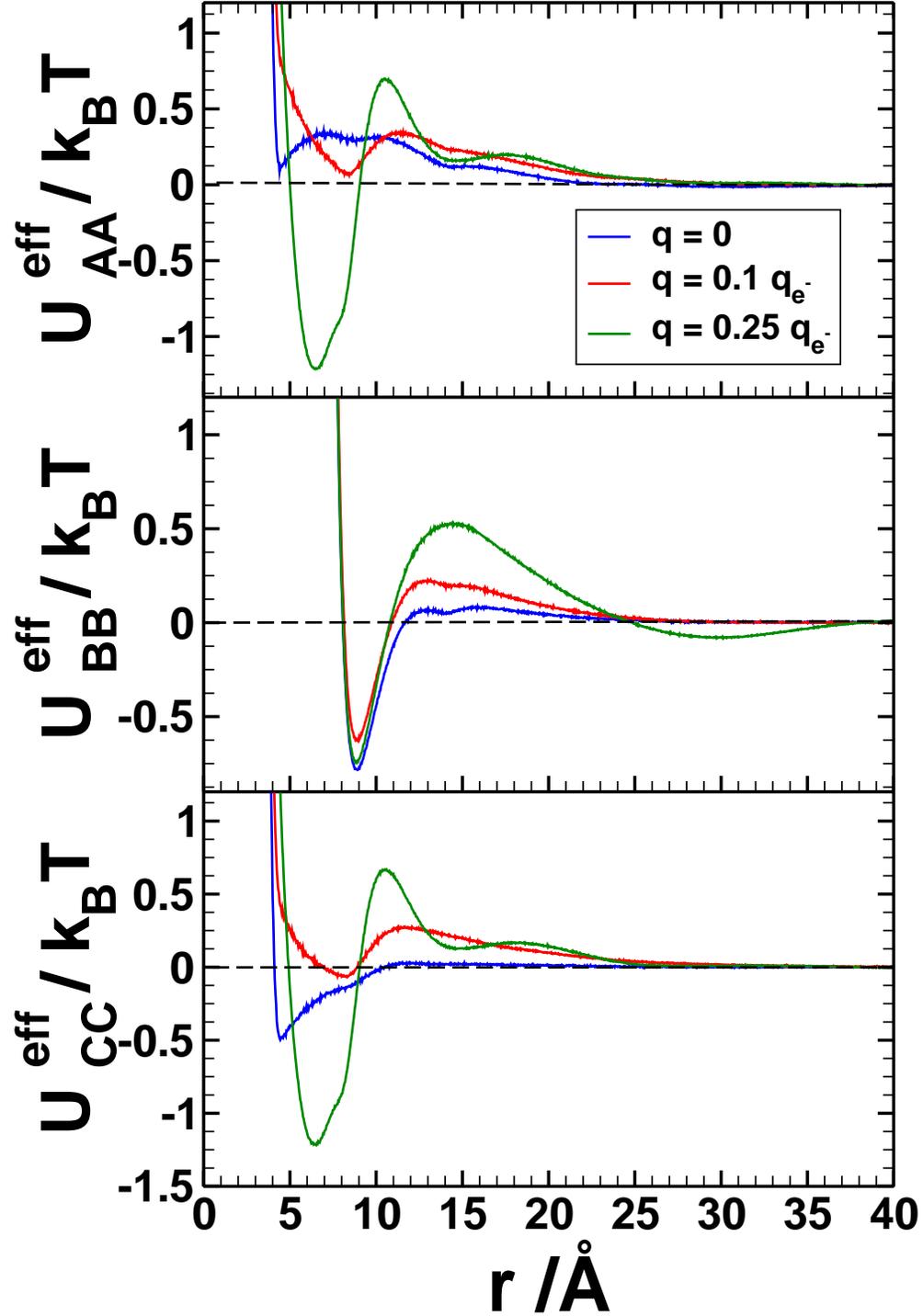}
\protect\protect\caption{Charge dependency of the effective potentials for A (top), B(middle)
  and C (bottom) particles. Charge magnitudes are specified in the legend. Values of $r_{cl}$ correspond to the inflexion points of the
effective potentials in their first minimum, i.e., $r_{cl}(A-A)=6$ \AA, $r_{cl}(B-B)=11$\AA, $r_{cl}(C-C)=10$\AA }

\label{fig:chargesEff} 
\end{figure}

\end{document}